\documentclass[aps,prl,onecolumn,groupedaddress]{revtex4}
\usepackage[dvips]{graphicx}
\usepackage{mathrsfs}
\usepackage{amsmath,amssymb}
\usepackage{hhline}
\usepackage{dcolumn}
\usepackage{bm}

\usepackage[dvips]{color}

\newcommand{\be}{\begin{equation}}
\newcommand{\ee}{\end{equation}}
\newcommand{\bi}{\begin{itemize}}
\newcommand{\ei}{\end{itemize}}

\begin{document}

\title{Emergence of Long-Range Correlations in Random Networks}
\date{\today}

\author{Shogo Mizutaka}
\email{mizutaka@jaist.ac.jp}
\affiliation{School of Knowledge Science, Japan Advanced Institute of Science and Technology, 1-1 Asahidai, Nomi 924-1292, Japan}
\author{Takehisa Hasegawa}
\email{takehisa.hasegawa.sci@vc.ibaraki.ac.jp}
\affiliation{Department of Mathematics and Informatics, Ibaraki University, 2-1-1 Bunkyo, Mito 310-8512, Japan}

\begin{abstract}
We perform an analytical analysis of the long-range degree correlation of the giant component in an uncorrelated random network by employing generating functions.
By introducing a characteristic length, we find that a pair of nodes in the giant component is negatively degree-correlated within the characteristic length and uncorrelated otherwise. 
At the critical point, where the giant component becomes fractal, the characteristic length diverges and the negative long-range degree correlation emerges.
We further propose a correlation function for degrees of the $l$-distant node pairs, which behaves as an exponentially decreasing function of distance in the off-critical region. The correlation function obeys a power-law with an exponential cutoff near the critical point.
The Erd\H{o}s-R\'{e}nyi random graph is employed to confirm this critical behavior. 
\end{abstract}
\maketitle

Most complex systems are described as networks comprising nodes and edges. Real network examples include cells, food webs, the Internet, the World Wide Web (WWW), social relationships, and companies' transactions \cite{caldarelli2007scale}.
Such real networks exhibit common structural properties, namely degree correlation, clustering, clique, motif, community structure, core-periphery structure, scale-free property, small-world property, and fractality \cite{caldarelli2007scale,newman2018networks,rombach2014core}.
Network science poses the fundamental question of how these properties relate to each other \cite{Chung2002,cohen2003scale,ravasz2003hierarchical,Vazquez2004,Stegehuis2017,xulvi2004reshuffling,Soffer2005,serrano2005tuning,radicchi2004defining,palla2005uncovering,arenas2008motif,fortunato2010community,orsini2015quantifying}.
In some networks, for example, small-world and fractal concept which are seemingly contradicting concepts coexist and they crossover from one to the other by varying the length scale \cite{kawasaki2010reciprocal,rozenfeld2010small}.
The small-world property represents an important attribute of real networks \cite{newman2018networks}.
In small-world networks, the average path length $\langle l\rangle$ between two nodes increases logarithmically with the system size $N$: $\langle l\rangle \sim \log N$, or equivalently $N\sim e^{{\rm const}\times\langle l\rangle}$.
Some real networks such as the WWW and protein interaction networks are fractal, as the number of boxes $N_{\rm B}(l)$ required to tile a network decreases with the increasing size $l$ of boxes according to a power-law: $N_{\rm B}(l)\sim l^{-d_{\rm B}}$, where $d_{\rm B}$ is the (finite) fractal dimension \cite{song2005self}.
By dividing the system size $N$ by the number of required boxes $N_{\rm B}(l)$, the average mass $\langle M_{\rm B}(l)\rangle = N/N_{\rm B}(l)$ of the boxes of size $l$ follows a power-law $\langle M_{\rm B}(l)\rangle \sim l^{d_{\rm B}}$.
This indicates that fractal dimension of the fractal networks is finite, while that of small-world networks becomes infinite, i.e., $d_{\rm B} \to \infty$.

In general, the fractal objects have no characteristic lengths--their structures are invariant under a length-scale transformation \cite{bunde2012fractals}.
A network is expected to have something invariant over a wide range of length scales when it is fractal.
Applying a renormalization technique to scale-free fractal networks demonstrates that the profiles of the degree distributions are invariant under renormalization \cite{song2005self}.
Negative degree correlation in scale-free fractal networks has been observed in various scales \cite{yook2005self}.
A scaling of the resistance and diffusion as a function of the distance and degrees of node pairs has been proposed \cite{gallos2007scaling}.
A scaling for degree correlations has been proposed under the assumption that the nearest-neighbor degree correlations of the fractal networks are invariant under renormalization \cite{gallos2008scaling}.
Previous studies \cite{song2005self,gallos2007scaling,gallos2008scaling,yook2005self} focusing on the structures and functions of the renormalized fractal networks have indicated that there is some correlation between its small- and large-scale network metrics, despite the difficulty in handling network renormalization.

With regard to local metrics of fractal networks, several reports addressed the correlation between the degrees of directly connected nodes by edges, i.e., the nearest-neighbor degree correlation.
Nearest-neighbor degree correlations are negative in various fractal networks, including empirical networks \cite{yook2005self}, synthetic networks \cite{fujiki2017fractality,song2006origins}, some trees \cite{goh2006skeleton,bialas2010long}, uncorrelated network models in a critical state \cite{bialas2008correlations,tishby2018revealing}, and percolating clusters of random networks \cite{mizutaka2018disassortativity} and clustered networks \cite{hasegawa2019structure}. 
(The converse is not true: the nearest-neighbor degree correlations do not make networks fractal \cite{fujiki2017fractality}).
Large-scale correlation structures of the fractal networks should be reflected in the degree correlation between nodes beyond their nearest-neighbors, i.e., the long-range degree correlation.
Fujiki {\it et al.} have introduced joint and associated conditional probabilities to analyze the long-range degree correlations of networks \cite{fujiki2018general}.
They have shown that in the large size limit, an uncorrelated random network satisfies the relation $P(k,k'|l)=q_{k}q_{k'}$, where $P(k,k'|l)$ is the probability that two randomly selected nodes separated by distance $l$ (two ends of a randomly selected $l$-chain) have degrees $k$ and $k'$, $q_{k}=kp_{k}/\langle k\rangle$ is the probability that an end of a randomly selected edge has $k$ edges, $p_{k}$ is a degree distribution, and $\langle k\rangle=\sum_{k}kp_{k}$.
A subsequent study pointed out that various networks, including fractal ones, exhibit long-range degree correlations \cite{fujiki2019identification}.
In \cite{rybski2010quantifying}, Rybski {\it et al.} have numerically analyzed the long-range degree correlations of the fractal networks described by the degree fluctuations in $l$-chains and indicated that the fractal networks have negative long-range correlations.
However, previous studies were performed numerically, and there are no analytical arguments for long-range degree correlations in the fractal networks.
In this study, we focus on the giant component of an uncorrelated random network. By characterizing its long-range degree correlation as a function of degrees of a pair of nodes and their distance, we analytically derive the emergence of the negative long-range degree correlation in the giant component at a critical state. 

Let us consider an infinitely-large uncorrelated random network with a degree distribution $p_{k}$ which has a locally tree-like structure.
The probability, $u$, that an edge does not lead to the giant component is given as the solution of $u=G_{1}(u)$, where $G_{1}(u)=\sum_{k}q_{k}u^{k-1}$, and it is less than $1$ in the presence of the giant component.
Assuming that a given network contains the giant component, i.e., $u<1$, we extract it from this network.
We start our analytical analysis by introducing probability $P_{{\rm GC}}(k,k'|l)$, stating that two ends of a randomly selected $l$-chain have degrees $k$ and $k'$, given that the chain belongs to the giant component.
Using the expectation number of nodes with degree $k'$ at distance $l$ from a degree-$k$ node and the expectation number of nodes at distance $l$ from a random node [see Eqs.~(18) and (21)], we obtain
\begin{align}
	P_{\rm GC}(k,k'|l)=\dfrac{ 1-v^{l-1}u^{k+k'-2} }{ 1-v^{l-1}u^{2} }q_{k}q_{k'}, \label{eq:pkkl}
	\end{align}
where $v$ is the probability that a node in between the $l$-chain is not connected to the giant component, which can be expressed as a function of $u$, as $v=G'_{1}(u)/G'_{1}(1)$ with $G'_{1}(u)=dG_{1}(u)/du=\sum_{k}(k-1)q_{k}u^{k-2}/\langle k\rangle$.
For $l=1$, it has been reported that $P_{\rm GC}(k,k'|l=1)=(1-u^{k+k'-2})q_{k}q_{k'}/(1-u^2)$, which depicts the joint probability that an edge selected randomly from the giant component is connected to degree-$k$ and -$k'$ nodes \cite{bialas2008correlations,tishby2018revealing}.

We introduce the characteristic lengths associated with the distance and the degrees of a node pair as
\begin{align}
	\xi_{l}=-1/\log{v}
	\label{eq:xi_l}
\end{align}
and
\begin{align}
	\xi_{k}=-1/\log{u},
	\label{eq:xi_k}
\end{align}
respectively.
Consequently, Eq.~(\ref{eq:pkkl}) is rewritten as
\begin{align}
		P_{\rm GC}(k,k'|l)=\dfrac{1 -e^{-(l-1)/\xi_{l}}e^{-(k+k'-2)/\xi_{k}}} {1-e^{-(l-1)/\xi_{l}}e^{-2/\xi_{k}} }q_{k}q_{k'}. \label{eq:pkkl_exp}
\end{align}
For a finite $\xi_{l}$, $P_{\rm GC}(k,k'|l)$ decays exponentially to $q_{k}q_{k'}$ with increasing chain length $l$.
Thus, a pair of $l$-distant nodes selected from the giant component is degree-correlated for $l\lesssim \xi_l$ and degree-uncorrelated for $l\gg \xi_l$.
Notably, relation $P(k,k'|l)=q_{k}q_{k'}$ holds for $l$-distant node pairs selected from the entire network.

We further introduce the probability, $P_{{\rm GC}}(k'|k,l)$, that one end of a chain has degree $k'$, given that the chain has a degree-$k$ node at the other end, has length $l$, and belongs to the giant component.
From Bayes' rule, this probability is given as 
\begin{align}
	P_{\rm GC}(k'|k,l)&=\frac{P_{\rm GC}(k,k'|l)}{\sum_{k'}P_{\rm GC}(k,k'|l)} \nonumber \\
	&= \dfrac{{1 -e^{-(l-1)/\xi_{l}}e^{-(k+k'-2)/\xi_{k}}}}
	{1-e^{-(l-1)/\xi_{l}}e^{-k/\xi_{k}}}q_{k'},
	\label{eq:pk_k'l}
\end{align}
and the average degree, $k_{l}^{\rm GC}(k)$, of $l$-distant nodes from degree-$k$ nodes on the giant component is given as
\begin{align}
	k_{l}^{{\rm GC}}(k)&=\sum_{k'}k'P_{{\rm GC}}(k'|k,l) \nonumber \\
	&=\dfrac{\langle k^2\rangle}{\langle k\rangle}+\dfrac{h(u)e^{-(l-1)/\xi_{l}}e^{-k/\xi_{k}}}
	{1-e^{-(l-1)/\xi_{l}}e^{-k/\xi_{k}}},
	\label{eq:kl_k}
\end{align}
where $h(u)=\sum_{k}kq_{k}(1-u^{k-2})\ge 0$.
Note that $k_{l}^{{\rm GC}}(k)$ of $l=1$ corresponds to the average degree of the nearest-neighbor degree-$k$ nodes on the giant component \cite{bialas2008correlations,tishby2018revealing}. 
\begin{figure*}[t!]
\begin{center}
\includegraphics[width=1.0\textwidth]{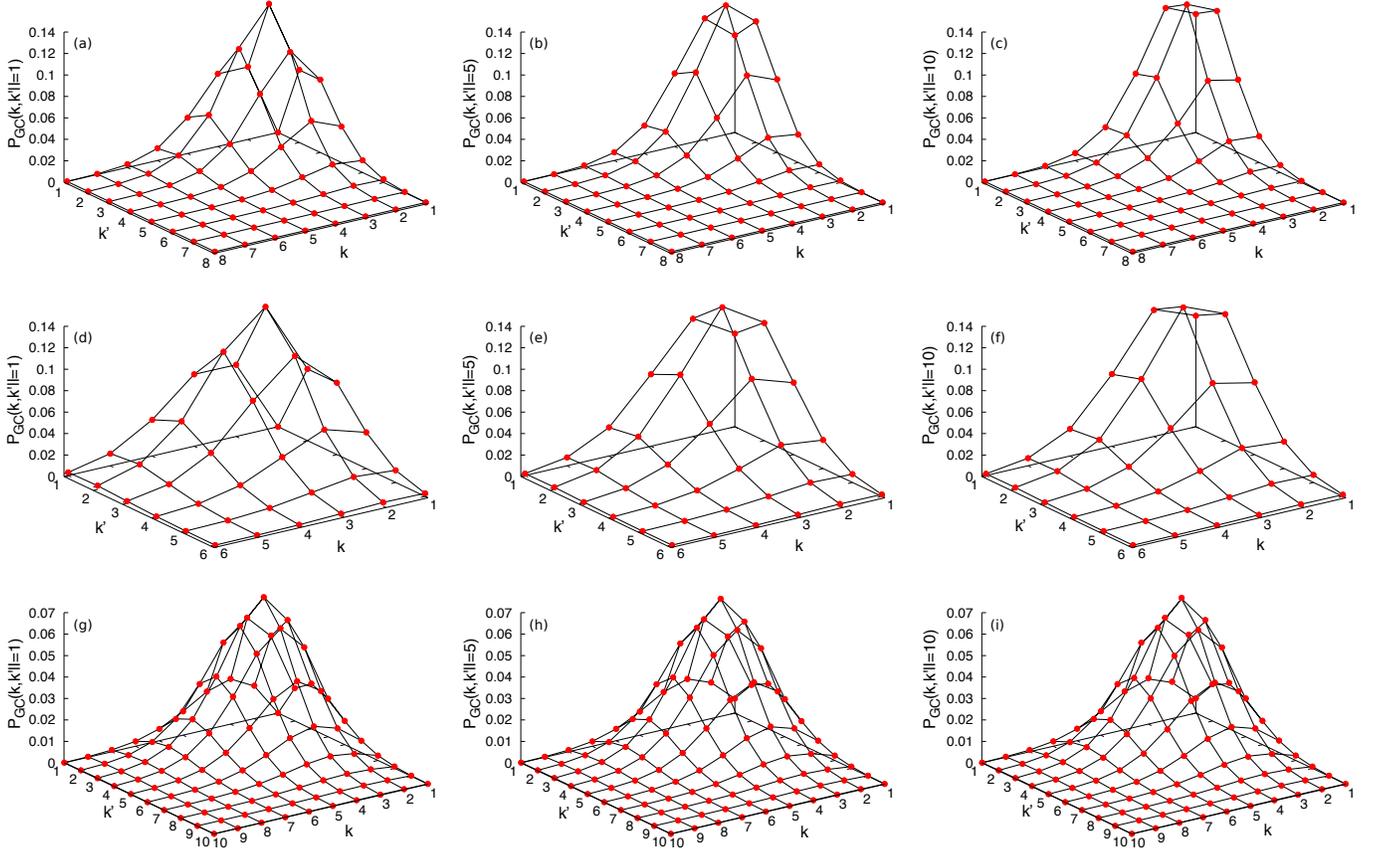}
\caption{
Probability distribution $P_{\rm GC}(k,k'|l)$ for Erd\H{o}s-R\'{e}nyi random graphs as a function of $k$ and $k'$ for several distances.
Wireframes depict the analytical calculation (\ref{eq:pkkl}), and symbols represent the corresponding simulation results.
Top panels represent the results for the average degree, $\lambda=1$, and distances (a) $l=1$, (b) $l=5$, and (c) $l=10$; middle panels represent the results for $\lambda=1.1$ and distance (d) $l=1$, (e) $l=5$, and (f) $l=10$; bottom panels depict the average degree, $\lambda=2.5$, and distance (g) $l=1$, (h) $l=5$, and (i) $l=10$. 
The results for each average degree are obtained from one sampled network of $10^7$ nodes.
}
\label{fig:pkkl}
\end{center}	
\end{figure*}
Equation~(\ref{eq:kl_k}) shows that $k_{l}^{\rm GC}(k)$ is a decreasing function of $k$ for a fixed value of $l$, indicating that the giant component is negatively degree-correlated. Moreover, $k_{l}^{\rm GC}(k)$ is a decreasing function of $l$ for any $k$, indicating that any degree correlation gradually disappears with increasing $l$ (as in Eq.~(\ref{eq:pkkl_exp})).
In summary, the giant component in a random network has a negative degree correlation for $l<\xi_{l}$, whereas it has no correlations for $l\gg\xi_{l}$: $k_{l}^{{\rm GC}}(k)\to \langle k^2\rangle/\langle k\rangle$. 


To further discuss the emergence of the long-range degree correlation of the giant component in detail, we employ the Erd\H{o}s-R\'{e}nyi random graph, whose degree distribution is $p_{k}=\lambda^{k}e^{-\lambda}/k!$, where $\lambda=\langle k \rangle$.
Prior to a detailed analysis, we test the validity of the theoretical analysis (\ref{eq:pkkl}) by comparing it with the simulation results.
In Fig.~\ref{fig:pkkl}, we plot the theoretical predictions (wireframes) of probability $P_{{\rm GC}}(k,k'|l)$ for the Erd\H{o}s-R\'{e}nyi random graphs with $\lambda=1(=\lambda_{\rm c})$, $\lambda=1.1$, and $\lambda=2.5$, where $\lambda_{\rm c}$ is the critical average degree above which the giant component exists.
The wireframes in all cases match the corresponding Monte-Carlo simulations (symbols) perfectly.

We assume that $\lambda=\lambda_{\rm c}+\delta$ and $u=v=1-\epsilon$, where both $\delta$ and $\epsilon$ are infinitely small values.
For $\lambda \gtrsim \lambda_{\rm c}$, we have $\epsilon\sim \lambda-\lambda_{\rm c}$ and two characteristic lengths, $\xi_{l}$ and $\xi_{k}$, as
\begin{align}
	\xi_{l}=\xi_{k} \sim \epsilon^{-1} \sim (\lambda-\lambda_{\rm c})^{-1}.
\end{align}
The critical exponents for $\xi_l$ and $\xi_k$ are unity, which corresponds to the critical exponent of the correlation (chemical) length for the mean size of the finite cluster in the percolation problem \cite{cohen2010complex}.   
For $\lambda \gtrsim \lambda_{\rm c}$, the second term of $k_{l}^{\rm GC}(k)$ becomes a power-law with an exponential cutoff of both $l$ and $k$ within $\xi_{l}$ and $\xi_{k}$ as
\begin{align}
	k_{l}^{\rm GC}(k)&=2+\frac{1}{(l-1)+k}e^{-(l-1)/\xi_{l}}e^{-(k-1)/\xi_{k}}.
		\label{eq:kl_k_errg_lc}
\end{align}
The two characteristic lengths in Eq.~(\ref{eq:kl_k_errg_lc}) diverge asymptotically in a critical state ($\lambda\to\lambda_{\rm c}$).
At $\lambda = \lambda_{\rm c}$, $k_{l}^{\rm GC}(k)$ decreases with increasing degree $k$ in a power-law for any $l(<\infty)$:
\begin{align}
	k_{l}^{\rm GC}(k)&= 2+\left(l+k-1\right)^{-1}.
		\label{eq:kl_k_errg_at_lc}
\end{align}
Hence, the negative long-range correlation in the giant component stretches entirely at criticality.

Furthermore, we propose {\it a degree-degree correlation function} $C(l)$, which characterizes the critical behavior of the networks.
Probability $P_{\rm GC}(k,k'|l)$ has full information on the structure of the giant component. 
We define the correlation function for degrees of $l$-distant node pairs on the GC, as $C(l)={\langle kk'\rangle}_{l}-{\langle k\rangle}_{l}{\langle k'\rangle}_{l}$, where ${\langle f(k,k')\rangle}_{l}=\sum_{k,k'}f(k,k')P_{\rm GC}(k,k'|l)$.
Combined with Eq.(\ref{eq:pkkl_exp}), $C(l)$ is expressed as 
\begin{align}
	C(l)&=\dfrac{-e^{-(l-1)/\xi_{l}}u^2\left(\sum_{k}kq_{k}\left(1-u^{k-2}\right)\right)^2
	}{\left(1 -e^{-(l-1)/\xi_{l}}u^{2}\right)^{2}}.
	\label{eq:Gl}
\end{align}
We observe that the degree correlation of $l$-distant node pairs on the giant component disappears for $l>\xi_{l}$, as $|C(l)|$ is an exponentially decreasing function of $l$.
The correlation function exhibits critical behavior when the giant component exists but infinitely small, i.e., $\xi_{l}\gg 1$: $C(l)$ in the critical region, and drops according to a power-law with an exponential cutoff,
\begin{align}
C(l) \sim-\frac{a^2}{(b(l-1)+2)^2}e^{-(l-1)/\xi_{l}},
	\label{eq:Gl_critical}
\end{align}
where $a=\langle k^2(k-2)\rangle/\langle k\rangle$ and $b=\langle k(k-1)(k-2)\rangle/\langle k\rangle$ \cite{assumption}.
At the critical point, $\xi_{l}$ diverges, and $C(l)\sim -l^{-2}$ for $l\gg 1$.
\begin{figure}[t!]
\begin{center}
\includegraphics[width=0.48\textwidth]{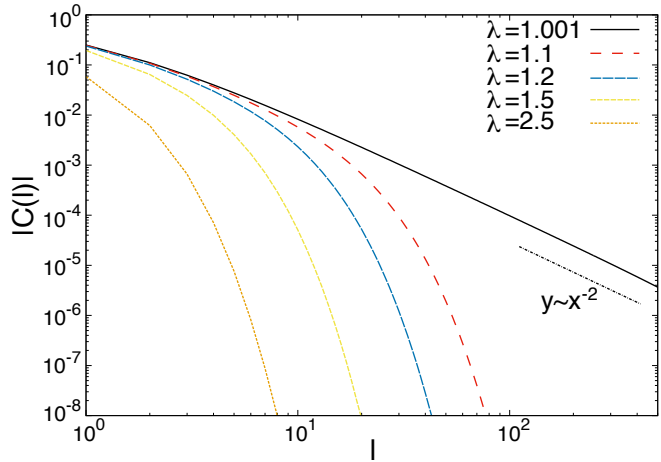}
\caption{
$|C(l)|$ of Erd\H{o}s-R\'{e}nyi random graphs as a function of $l$.
Lines from right to left correspond to $|C(l)|$ for $\lambda=1.001$, $1.1$, $1.2$, $1.5$, and $2.5$, respectively.
Dashed line with slope $-2$ is plotted as a guide to the eye.
}
\label{fig:gl}
\end{center}	
\end{figure}
Figure~\ref{fig:gl} shows the absolute value $|C(l)|$ of the correlation function for Erd\H{o}s-R\'{e}nyi random graphs for several values of $\lambda$.
We observe that the correlation function decays exponentially in the off-critical region ($\lambda\ge 1.1$), and a power-law with exponent $-2$ exists near criticality ($\lambda=1.001\approx\lambda_{\rm c}$). 

All our analyses conclude that the long-range degree correlation in the giant component of an uncorrelated random network emerges at the critical point.
Both $P_{\rm GC}(k,k'|l)$ and $k_{l}^{\rm GC}(k)$ indicate that the giant component of an uncorrelated random network exhibits a long-range degree correlation. The giant component is negatively correlated for $l< \xi_{l}$, whereas it becomes neutral for $l\gg \xi_{l}$.
At criticality, where $\xi_{l}$ diverges and the giant component is fractal, the negative degree correlation is observed at any distance.
Moreover, the correlation function $C(l)$ for degrees of $l$-distant node pairs decays exponentially in the off-critical region.
In contrast, it obeys a power-law with a cutoff, $C(l)\sim -l^{-2}e^{l/\xi_{l}}$ for $l\gg 1$ near criticality and becomes a power-law, $C(l)\sim -l^{-2}$ at criticality.
In summary, the negative long-range degree correlation spontaneously emerges in the fractal networks. 

The long-range degree correlation for a given network is intrinsic or extrinsic.
Extrinsic ones are correlations arisen form nearest-neighbor degree correlations.
We note that even when a given network has only a strong negative nearest-neighbor degree correlation, it will exhibit an extrinsic long-range degree correlation, such that the degrees of node pairs are positively correlated at $l=2,4,\cdots$.
Extrinsic correlations can be described by the products of probability $P(k'|k)$ that a random neighbor of a degree-$k$ node has $k'$ edges.
In an extrinsic case, for example, a triplet probability, $P(k',k''|k)$,  that a degree-$k$ node is connected to a degree-$k'$ node and a degree-$k''$ node satisfies $P(k',k''|k)=P(k'|k)P(k''|k)$.
As stated in \cite{fujiki2019identification}, most empirical networks have intrinsic correlations which cannot be explained by only nearest-neighbor degree correlations.
With regard to the giant component in an uncorrelated random network, the long-range degree correlation is considered intrinsic. The triplet probability $P_{\rm GC} (k',k''|k)$ indicates $P_{\rm GC}(k',k''|k)=(1-u^{k+k'+k''-4})q_{k'}q_{k''}/(1-u^{k})\neq P_{\rm {GC}}(k'|k)P_{\rm {GC}}(k''|k)$.

Bialas and Ole\'s have investigated the correlation function for generic trees \cite{bialas2010long}. Their correlation function (Eq.~(12) in \cite{bialas2010long}) behaves as a power-law, which is similar to $C(l)$ in the present study.
Interestingly, the long-range degree correlation of the giant component is attributed to the divergence of the correlation length in the phase transition, while the behavior in the generic trees is not associated with criticality. (However, we note that $\xi_{l}$ is directly connected to the correlation length through Eq.~(2.29) in \cite{bunde2012fractals}.)
The tree structures found in both generic trees and the giant component at criticality may result in a power-law behavior of $C(l)$, although further studies are required to gain an understanding of this common mechanism.

Previous works have captured nearest-neighbor degree correlations of fractal networks which are renormalized at several length scales, implying a correlation between small- and large-scale degree correlations \cite{gallos2008scaling,yook2005self}.
We did not attain the problem how the long-range degree correlation treated in this study associates with the nearest-neighbor degree correlations of renormalized networks.
This study deepens our understanding of the relation between the emergence of the long-range degree correlation and criticality/fractality.

\textit{Acknowledgement.}
The authors would like to thank K. Yakubo for fruitful discussions.
S.M.\ and T.H.\ acknowledge the financial support from JSPS (Japan) KAKENHI Grant Number JP18KT0059.
S.M.\ was supported by a grant-in-aid for Early-Career Scientists (No.~18K13473) and a grant-in-aid for JSPS Research Fellow (No.~18J00527) from the Japan Society for the Promotion of Science (JSPS).
T.H.\ acknowledges the financial support from JSPS (Japan) KAKENHI, grant number JP19K03648.

\bibliography{ref.bib}

%
\begin{widetext}

\section{Derivation of $P_{\rm GC}(k_{\rm s},k_{\rm t})$}

Let us consider an infinitely-large uncorrelated random network with a locally tree-like structure and focus on $l$-distant nodes, which are nodes at distance $l$ from a randomly chosen node (seed).
We suppose that for a randomly chosen seed the numbers of $l$-distant nodes with degrees $1,2,\cdots,k,\cdots,$ and $K$ are $n_1,n_2,\cdots,n_k,\cdots,$ and $n_K$, respectively, where $K$ is the maximum degree. 
We denote the sequence of the number $l$-distant nodes with each degree by ${\bm n}=(n_1,n_2,\cdots,n_K)$.
Let $P_{l}({\bm n},{\rm GC}|N_l)$ be the probability that the sequence for the $l$-distant nodes is ${\bm n}$ and the focal component belongs to the giant component under the condition that the total number of $l$-distant nodes $N_l(=\sum_{k=1}^{K} n_k)$ is given.
Using the multinomial distribution, $P_{l}({\bm n},{\rm GC}|N_l)$ is  given as
\begin{align}
P_{l}({\bm n},{\rm GC}|N_l)=\frac{N_{l}!}{n_{1}!n_{2}!\cdots n_{k}!\cdots n_{K}!}\left(1-\prod_{k=1}^{K}(u^{k-1})^{n_{k}}\right)\prod_{k=1}^{K} q_{k}^{n_k},
\end{align}
where $u$ is the probability that an edge does not lead to the giant component, $p_k$ is the degree distribution, and $q_k =\sum_{k}kp_{k}/\langle k\rangle$ is the neighbor's degree distribution.
The term $ 1-\prod_{k=1}^{K}(u^{k-1})^{n_{k}}$ corresponds with the probability that at least one of outgoing edges from $l$-distant nodes connects to the giant component.
By combining $P_l({\bm n},{\rm GC}|N_l)$ and $P_{l}(N_{l}|k_{\rm s})$, which is the probability that the number of $l$-distant nodes is $N_{l}$ given that the seed has $k_{\rm s}$ edges,
we construct the probability $P({\bm n},{\rm GC}|k_{\rm s})$ that the sequence for the $l$-distant nodes is ${\bm n}$ and the focal component belongs to the giant component given that the seed has $k_{\rm s}$ edges as
\begin{align}
P_{l}({\bm n},{\rm GC}|k_{\rm s})&=\sum_{N_l}P_{l}(N_{l}|k_{\rm s})P_l({\bm n},{\rm GC}|N_l) \nonumber \\
	&=\sum_{N_l}P_{l}(N_{l}|k_{\rm s})\frac{N_{l}!}{n_{1}!n_{2}!\cdots n_{k}!\cdots n_{K}!}\left(1-\prod_{k=1}^{K}(u^{k-1})^{n_{k}}\right)\prod_{k=1}^{K} q_{k}^{n_k}.
\end{align}
We introduce the generating function $\tilde{F}_{\rm GC}({\bm x} | k_{\rm s})$ for $P_{l}({\bm n},{\rm GC}|k_{\rm s})$ as
\begin{equation}
\tilde{F}_{\rm GC}({\bm x} | k_{\rm s})=\sum_{\bm n} P_{l}({\bm n},{\rm GC}|k_{\rm s}) \prod_{k=1}^K x_k^{n_k}, 
\;\, {\rm where} \;\, {\bm x}=(x_1,x_2,\cdots,x_K).
\end{equation}
This function is calculated as follows:
\begin{align}
	\tilde{F}_{\rm GC}({\bm x} | k_{\rm s})
	&=\sum_{{\bm n}}\sum_{N_l}P_{l}(N_{l}|k_{\rm s})\frac{N_{l}!}{n_{1}!n_{2}!\cdots n_{k}!\cdots n_{K}!}\prod_{k=1}^{K} q_{k}^{n_k}x_{k}^{n_k}\left(1-\prod_{k=1}^{K}(u^{k-1})^{n_{k}}\right)\prod_{k=1}^{K} q_{k}^{n_k}x_{k}^{n_k} \nonumber \\
	&=\sum_{N_l}\sum_{{\bm n}}P_{l}(N_{l}|k_{\rm s})\frac{N_{l}!}{n_{1}!n_{2}!\cdots n_{k}!\cdots n_{K}!}\left(\prod_{k=1}^{K} (q_{k}x_{k})^{n_{k}}-\prod_{k=1}^{K} (q_{k}x_{k}u^{k-1})^{n_{k}}\right) \nonumber \\
	&=\sum_{N_l}P_{l}(N_{l}|k_{\rm s})\left(\left(\sum_{k=1}^{K}q_{k}x_{k}\right)^{N_{l}}-\left(\sum_{k=1}^{K}q_{k}x_{k}u^{k-1}\right)^{N_{l}}\right) \nonumber \\
	&=\tilde{G}_{l}\left(\textstyle{\sum_{k}} q_{k}x_{k} | k_{\rm s}\right)-\tilde{G}_{l}\left(\textstyle{\sum_{k}} q_{k}x_{k}u^{k-1} | k_{\rm s}\right).	\label{eq:ftilde}
\end{align}
Here $\tilde{G}_{l}(x | k_{\rm s})$ is the generating function for the probability distribution $P_{l}(N_{l}|k_{\rm s})$ and is given as
\begin{align}
	\tilde{G}_{l}(x | k_{\rm s})=(G_{1}(G_{1}(\cdots(G_{1}(x))\cdots)))^{k_{\rm s}},
\end{align}
which is known as the generating function for the distribution of the number of children in $l$ generation under a given offspring distribution $q_k$ and initial population $k_{s}$ \cite{harris1964theory}.
The probability distribution $P_{l}({\bm n},{\rm GC},k_{\rm s})$ that the sequence for the $l$-distant nodes is ${\bm n}$, the seed has $k_{\rm s}$ edges, and the focal component belongs to the giant component is $P_{l}({\bm n},{\rm GC},k_{\rm s})=P_{l}({\bm n},{\rm GC}|k_{\rm s})p_{k_{\rm s}}$, and the corresponding generating function $\tilde{F}_{\rm GC}({\bm x}, k_{\rm s})(=\sum_{\bm n} P_{l}({\bm n},{\rm GC},k_{\rm s}) \prod_{k=1}^K x_k^{n_k})$ is thus given by the product of $\tilde{F}_{\rm GC}({\bm x}|k_{\rm s})$ and $p_{k_{\rm s}}$,
\begin{align}
F_{\rm GC}({\bm x}, k_{\rm s})
&=p_{k_{\rm s}}\tilde{F}_{\rm GC}({\bm x}|k_{\rm s}) \nonumber \\
&=p_{k_{\rm s}}\tilde{G}_l\left(\textstyle{\sum_{k}}q_{k}x_{k} |k_{\rm s}\right)-p_{k_{\rm s}}\tilde{G}_l\left(\textstyle{\sum_{k}}q_{k}x_{k}u^{k-1}|k_{\rm s}\right).
\label{eq:gf_from_ks}	
\end{align}

Differentiating Eq.~(\ref{eq:gf_from_ks}) with respect to $x_{k_{\rm t}}$ and substituting ${\bm x}={\bold 1}$ into it, we have the expectation number $\langle N^{k_{\rm s},k_{\rm t}}(l,{\rm GC})\rangle$ of $l$-distant nodes with degree $k_{\rm t}$ where a randomly chosen seed has $k_{\rm s}$ edges and its component belongs to the giant component:
\begin{align}
	\langle N^{k_{\rm s},k_{\rm t}}(l,{\rm GC})\rangle
	&=\frac{\partial F_{\rm GC}({\bm x}, k_{\rm s})}{\partial x_{k_{\rm t}}}\biggr|_{{\bm x}={\bf 1}} \nonumber \\
	&=k_{\rm s}p_{k_{\rm s}}q_{k_{\rm t}}G'^{l-1}_{1}(1)\left(1-v^{l-1}u^{k_{\rm s}+k_{\rm t}-2}\right),
	\label{eq:exptation_number_kk}
\end{align}
where $v=G'_{1}(u)/G'_{1}(1)$.

In a similar way, we easily find the generating function $F_{\rm GC}({\bm x})$ for the probability distribution $P_{l}({\bm n},{\rm GC})$ that for a randomly chosen node, the sequence of the number of $l$-distant nodes with each degree is ${\bm n}$ and these nodes are member of the giant component as
\begin{align}
	F_{\rm GC}({\bm x})=\sum_{k_{\rm s}=1}^K F_{\rm GC}({\bm x},k_{\rm s})
	=G_l\left(\textstyle{\sum_{k}} q_{k}x_{k}\right)-G_l\left(\textstyle{\sum_{k}} q_{k}x_{k}u^{k-1} \right),
	\label{eq:gf_from_random_to_random}
\end{align}
where
\begin{align}
	G_l(x)&=\sum_{k}p_{k}\tilde{G}_l(x|k)\nonumber \\
	&=G_{0}(G_{1}(G_{1}(\cdots(G_{1}(x))\cdots)))
\end{align}
From Eq.~(\ref{eq:gf_from_random_to_random}), the expectation number $\langle N(l,{\rm GC})\rangle$ of $l$-distant nodes from a random seed which belong to the giant component is calculated as
\begin{align}
	\langle N(l,{\rm GC})\rangle
	&=\sum_{k_{\rm t}=1}^K \frac{\partial F_{\rm GC}({\bm x})}{\partial x_{k_{\rm t}}}\biggr|_{{\bm x}={\bm 1}} \nonumber \\
	&=\langle k\rangle G'^{l-1}_{1}(1)\left(1-v^{l-1}u^{2}\right)
	\label{eq:expctation_number}
\end{align}
By dividing Eq.~(\ref{eq:exptation_number_kk}) by Eq.~(\ref{eq:expctation_number}), we obtain the probability $P_{\rm GC}(k_{\rm s},k_{\rm t}|l)$ that two ends of a randomly chosen $l$-chain from the giant component have degree $k_{\rm s}$ and $k_{\rm t}$ as
\begin{align}
	P_{\rm GC}(k_{\rm s},k_{\rm t}|l)&\equiv \frac{\langle N^{k_{\rm s},k_{\rm t}}(l,{\rm GC})\rangle}{\langle N(l,{\rm GC})\rangle} \nonumber \\
	&=\dfrac{G'_{1}(1)^{l-1}\left( 1-v^{l-1}u^{k_{\rm s}+k_{\rm t}-2} \right)}{ G'_{1}(1)^{l-1}\left( 1-v^{l-1}u^{2} \right)}q_{k_{\rm s}}q_{k_{\rm t}}, \nonumber \\
	&=\dfrac{ 1-v^{l-1}u^{k_{\rm s}+k_{\rm t}-2}}{ 1-v^{l-1}u^{2} }q_{k_{\rm s}}q_{k_{\rm t}},
	\label{eq:pkk'l}
\end{align}
where we call a connected path with length $l$ as an $l$-chain.
Here, the denominator is proportional to the number of $l$-chains in the giant component in that $\langle k\rangle G'_{1}(1)^{l-1}$ ($\langle k\rangle G'_{1}(1)^{l-1}v^{l-1}u^2$) represents the average number of nodes at distance $l$ from a randomly chosen node in the whole network (finite components).
When the networks are singly connected, i.e., $u=0$, Eq.~(\ref{eq:pkk'l}) reduces to $P(k_{\rm s},k_{\rm t}|l)=q_{k_{\rm s}}q_{k_{\rm t}}$ which is a known result for uncorrelated random networks \cite{fujiki2018general}.
Taking the limit $u\to 1$ ($v\to 1$), we obtain $P_{\rm GC}(k_{\rm s},k_{\rm t}|l)$ at criticality as
\begin{align}
	\lim_{u\to 1, v\to 1} P_{\rm GC}(k_{\rm s},k_{\rm t}|l,{\rm GC})	=\dfrac{\dfrac{G''_{1}(1)}{G'_{1}(1)}(l-1)+(k_{\rm s}+k_{\rm t}-2)}{\dfrac{G''_{1}(1)}{G'_{1}(1)}(l-1)+2}q_{k_{\rm s}}q_{k_{\rm t}}.
	\label{eq:pkk'l_pt}
\end{align}
Replacing $u$ and $v$ by $\xi_{k}=-1/\log{u}$ and $\xi_{l}=-1/\log{v}$, respectively, we can rewrite Eq.~(\ref{eq:pkk'l}) as
\begin{align}
	P_{\rm GC}(k_{\rm s},k_{\rm t}|l)
	&=\dfrac{1 -e^{-(l-1)/\xi_{l}}e^{-(k_{\rm s}+k_{\rm t}-2)/\xi_{k}}} {1-e^{-(l-1)/\xi_{l}}e^{-2/\xi_{k}} }q_{k_{\rm s}}q_{k_{\rm t}}.
	\label{eq:other_pkk'l2}
\end{align}

Let us introduce the probability $P(k_{\rm t}|k_{\rm s},l,{\rm GC})$ that one end of a chain has degree $k_{\rm t}$ given that the chain has length $l$ and has a degree-$k_{\rm s}$ node as a starting node, and belongs to the giant component. The probability is given from Eq.~(\ref{eq:pkk'l}) as
\begin{align}
	P(k_{\rm t}|k_{\rm s},l,{\rm GC})
	&=\dfrac{1-v^{l-1}u^{k_{\rm s}+k_{\rm t}-2} }{ 1-v^{l-1}u^{k_{\rm s}}}q_{k_{\rm t}}.
	\label{eq:pk_k'l2}
\end{align}
We get the average degree $k_{l}^{\rm GC}(k_{\rm s})$ of $l$-distant nodes from a degree-$k_{\rm s}$ node on giant component as
\begin{align}
	k_{l}^{\rm GC}(k_{\rm s})&=\sum_{k_{\rm t}}k_{\rm t}P(k_{\rm t}|k_{\rm s},l,{\rm GC}) \nonumber \\
	&=\frac{\langle k^2\rangle}{\langle k\rangle}+\frac{h(u)v^{l-1}u^{k_{\rm s}}}{1-v^{l-1}u^{k_{\rm s}}}
	\label{eq:kl_GC_gne}
\end{align}
where $h(u)=\sum_{k}kq_{k}(1-u^{k-2})$.
We consider the situation that the giant component exists but infinitely small i.e., $u\sim 1-\epsilon$ and $v\sim 1- \langle k(k-1)(k-2)\rangle\epsilon/\langle k(k-1)\rangle$. In the situation, Eq.~(\ref{eq:kl_GC_gne}) approximates  
\begin{align}
	k_{l}^{\rm GC}(k_{\rm s})&\sim \frac{\langle k^2\rangle}{\langle k\rangle}\left(1+\frac{\langle k^3\rangle-2\langle k^2\rangle}{\langle k^2\rangle}\frac{1}{(l-1)\langle k(k-1)(k-2)\rangle/\langle k(k-1)\rangle+k_{\rm s}}\right).
	\label{kl_k_near}
\end{align}

\section{Correlation function $C(l)$}

We consider the correlation function $C(l)$ defined as
\begin{align}
	C(l)={\langle kk'\rangle}_{l}-{\langle k\rangle}_{l}{\langle k'\rangle}_{l},
	\label{eq:cl}
\end{align}
where ${\langle f(k,k')\rangle}_{l}=\sum_{k,k'}f(k,k')P_{\rm GC}(k,k'|l)$.
From Eqs.~(\ref{eq:pkk'l}) and (\ref{eq:cl}), we have
\begin{align}
	C(l)&=\sum_{k,k'}kk'\dfrac{1 -v^{l-1}u^{k+k'-2}}{1 -v^{l-1}u^{2} }q_{k}q_{k'}-\left(\sum_{k,k'}k\dfrac{1 -v^{l-1}u^{k+k'-2}}{1 -v^{l-1}u^{2} }q_{k}q_{k'}\right)^{2} \nonumber\\
	&=\dfrac{\left(\frac{\langle k^2\rangle}{\langle k\rangle}\right)^{2} -v^{l-1}\left(\sum_{k}kq_{k}u^{k-1}\right)^{2}}{1 -v^{l-1}u^{2} }-\left(\dfrac{\frac{\langle k^2\rangle}{\langle k\rangle} -v^{l-1}\left(\sum_{k}kq_{k}u^{k-1}\right)u}{1 -v^{l-1}u^{2} }\right)^{2}.
	\label{eq:correlation_function}
\end{align}
The first term of the right hand side in Eq.~(\ref{eq:correlation_function}) is
\begin{align}
	&\dfrac{\left(\frac{\langle k^2\rangle}{\langle k\rangle}\right)^{2} -v^{l-1}\left(\sum_{k}kq_{k}u^{k-1}\right)^{2}}{1 -v^{l-1}u^{2} }\nonumber \\
	&=\dfrac{\left({1 -v^{l-1}u^{2} }\right)\left(\left(\frac{\langle k^2\rangle}{\langle k\rangle}\right)^{2} -v^{l-1}\left(\sum_{k}kq_{k}u^{k-1}\right)^{2}\right)}{\left(1 -v^{l-1}u^{2}\right)^{2} } \nonumber \\
	&=\dfrac{\left(\frac{\langle k^2\rangle}{\langle k\rangle}\right)^{2}-\left(\frac{\langle k^2\rangle}{\langle k\rangle}\right)^{2}v^{l-1}u^{2}+v^{2(l-1)}\left(\sum_{k}kq_{k}u^{k-1}\right)^{2}u^2-v^{l-1}\left(\sum_{k}kq_{k}u^{k-1}\right)^{2}
	}{\left(1 -v^{l-1}u^{2}\right)^{2} },
	\label{eq:first_term}
\end{align}
and the second term is
\begin{align}
	&\left(\dfrac{\frac{\langle k^2\rangle}{\langle k\rangle} -v^{l-1}\left(\sum_{k}kq_{k}u^{k-1}\right)u}{1 -v^{l-1}u^{2} }\right)^{2} \nonumber \\
	&=\frac{\left(\frac{\langle k^2\rangle}{\langle k\rangle}\right)^{2}-2\frac{\langle k^2\rangle}{\langle k\rangle}v^{l-1}\left(\sum_{k}kq_{k}u^{k-1}\right)u+v^{2(l-1)}\left(\sum_{k}kq_{k}u^{k-1}\right)^{2}u^{2}}{(1 -v^{l-1}u^{2})^{2} }.
	\label{eq:second_term}
\end{align}

Then, we have
\begin{align}
	C(l)&=\dfrac{-v^{l-1}\left(\sum_{k}kq_{k}u^{k-1}\right)^{2}
	-\left(\frac{\langle k^2\rangle}{\langle k\rangle}\right)^{2}v^{l-1}u^{2}+2\frac{\langle k^2\rangle}{\langle k\rangle}v^{l-1}\left(\sum_{k}kq_{k}u^{k}\right)
	}{\left(1 -v^{l-1}u^{2}\right)^{2}} \nonumber \\
	&=\dfrac{-e^{-(l-1)/\xi_{l}}\left(\sum_{k}kq_{k}u^{k-1}-u\frac{\langle k^2\rangle}{\langle k\rangle}\right)^2
	}{\left(1 -e^{-(l-1)/\xi_{l}}u^{2}\right)^{2}},
	\label{eq:Gl_SM}
\end{align}
where we used $v=e^{-1/\xi_{l}}$.
Expanding $u$ and $e^{-(l-1)/\xi_{l}}$ in the denominator as $u\sim1-\epsilon$ and $e^{-(l-1)/\xi_{l}}\sim 1-(l-1)/\xi_{l}$, we drive the correlation function in the critical region as
\begin{align}
	C(l)&\sim -\frac{a^2}{(b(l-1)+2)^2}e^{-(l-1)/\xi_{l}},
\end{align}
where $a=\langle k^2(k-2)\rangle/\langle k\rangle$ and $b=\langle k(k-1)(k-2)\rangle/\langle k\rangle$.	
\end{widetext}

\end{document}